\documentclass[aps,nofootinbib,nobibnotes,amsmath,amssymb,amsfonts,noshowpacs,floatfix]{revtex4-1}

\usepackage{graphicx,bm,epsfig,setspace}
\usepackage{color}
\usepackage{hyperref}
\usepackage{epstopdf}
\usepackage{ulem}
\usepackage{soul}
\usepackage{xparse}
\usepackage{mathtools}
\usepackage{subfig}

\usepackage[labelfont=bf, textfont=small, 
            justification=justified,
            singlelinecheck=false,format=plain]{caption}

\makeatletter
\def\graphicscale{\twocolumn@sw{0.33}{0.4}}
\makeatother

\ExplSyntaxOn
\NewDocumentCommand{\mref}{m}{\quinn_mref:n {#1}}
\seq_new:N \l_quinn_mref_seq
\cs_new:Npn \quinn_mref:n #1
 {
  \seq_set_split:Nnn \l_quinn_mref_seq { , } { #1 }
  \seq_pop_right:NN \l_quinn_mref_seq \l_tmpa_tl
  ( 
  \seq_map_inline:Nn \l_quinn_mref_seq
    { \ref{##1},\nobreakspace } 
  \exp_args:NV \ref \l_tmpa_tl 
  ) 
 }
\ExplSyntaxOff

\def\spose#1{\hbox to 0pt{#1\hss}}
\def\lesssim{\mathrel{\spose{\lower 3pt\hbox{$\mathchar"218$}}
 \raise 2.0pt\hbox{$\mathchar"13C$}}}
\def\gtrsim{\mathrel{\spose{\lower 3pt\hbox{$\mathchar"218$}}
 \raise 2.0pt\hbox{$\mathchar"13E$}}}
\def\
case#1#2{{\textstyle\frac{#1}{#2}}}
\def\<{\langle}
\def\>{\rangle}
 \def\({\left(}
\def\){\right)}
 \def\[{\left[}
\def\]{\right]}

\DeclareMathOperator\erfc{erfc}
\newcommand*{\beq}{\begin{equation}}
\newcommand*{\eeq}{\end{equation}}
\newcommand*{\bea}{\begin{eqnarray}}
\newcommand*{\eea}{\end{eqnarray}}

\newcommand \hrho{\hat{\rho}_1}

\definecolor{ao(english)}{rgb}{0.0, 0.5, 0.0}
\definecolor{forestgreen(web)}{rgb}{0.13, 0.55, 0.13}

\newcommand{\isEquivTo}[1]{\underset{#1}{\sim}}

\DeclareMathOperator{\integer}{int}

\def\simge{\mathrel{%
       \rlap{\raise 0.511ex \hbox{$>$}}{\lower 0.511ex \hbox{$\sim$}}}}
\def\simle{\mathrel{
       \rlap{\raise 0.511ex \hbox{$<$}}{\lower 0.511ex \hbox{$\sim$}}}}

\begin{document}

\title{On the force--velocity relationship of a bundle of rigid living filaments.}

\author{Alessia Perilli}
\email{alessia.perilli@roma1.infn.it}
\affiliation{Department of Physics, Sapienza University of Rome, P.le Aldo Moro 5, I-00185 Rome, Italy;\\ and, Department of Chemistry, \'Ecole Normale Superi\'eure, rue Lhomond 24, 75005 Paris}
\author{Carlo Pierleoni}
\email{carlo.pierleoni@aquila.infn.it}
\affiliation{Department of Physical and Chemical Sciences, University of L'Aquila, Via Vetoio 10, 67100  L'Aquila, Italy;\\ and, Maison de la Simulation, CEA--Saclay 91120 Gif sur Yvette, France}
\author{Giovanni Ciccotti}
\email{giovanni.ciccotti@roma1.infn.it}
\affiliation{Instituto per le Applicazioni del Calcolo ``Mauro Picone'' (IAC), CNR, Via dei Taurini 19, I-00185 Rome, Italy;\\ and, Sapienza University of Rome, P.le Aldo Moro 5, I-00185 Rome, Italy; and, University College Dublin (UCD), Belfield, Dublin 4, Ireland}
\author{Jean-Paul Ryckaert}
\email{jryckaer@ulb.ac.be}
\affiliation{Department of Physics, Universit\'e Libre de Brussels (ULB), Campus Plaine, CP 223, 
B-1050 Brussels, Belgium}

\date{\today}

\begin{abstract}
In various cellular processes, biofilaments like F--actin and F--tubulin are able to exploit chemical energy associated to polymerization to perform mechanical work against an external load. 
The force--velocity relationship quantitatively summarizes the nature of this process.  
By a stochastic dynamical model, we give, together with the evolution of a staggered bundle of $N_f$ rigid living filaments facing a loaded wall, the corresponding force--velocity relationship.
We compute systematically the simplified evolution of the model in supercritical conditions $\hrho=U_0/W_0>1$ at $\epsilon=d^2W_0/D=0$, where $d$ is the monomer size, $D$ is the obstacle diffusion coefficient, $U_0$ and $W_0$ are the polymerization and depolymerization rates.
Moreover, we see that the solution at $\epsilon=0$ is valid for a good range of small non--zero $\epsilon$ values.
We consider two classical protocols: the bundle is opposed either to a constant load or to an optical trap set--up, characterized by a harmonic restoring force. 
The constant force case leads, for each $F$ value, to a stationary velocity $V^{stat}(F;N_f,\rho_1)$ after a relaxation with characteristic time $\tau_{micro}(F)$. 
When the bundle (initially taken as an assembly of filament seeds) is subjected to a harmonic restoring force (optical trap load), the bundle elongates and the load increases up to stalling (equilibrium) over a characteristic time $\tau^{OT}$.
Extracted from this single experiment, the force--velocity $V^{OT}(F;N_f,\rho_1)$ curve is found to coincide with $V^{stat}(F;N_f,\rho_1)$, except at low loads.
We show that this result follows from the adiabatic separation between $\tau_{micro}$ and $\tau^{OT}$, i.e. $\tau_{micro}\ll\tau^{OT}$.
\\
\\
\textit{Submitted to the Journal of Chemical Physics on August 22nd 2017}
\end{abstract}
\maketitle

\section{Introduction}

Cell motility \textit{in vivo} is a large scale manifestation of the living character of the cytoskeleton bio--filaments network \cite{Risler2009}. In particular, F--actin filaments produce growing lamellipodium or filopodium structures where G--actin monomers polymerize at the barbed end of filaments, directly in contact with the cytoplasmic membrane. 
The speed of the membrane deformation/displacement at the leading edge of the cell adjusts itself so that the force generated by the growing filaments compensates the resisting load 
coming from the membrane tension and to the crowded environment around the cell. 
For living filaments opposing a loaded mobile obstacle, the macroscopic force--velocity relationship, $V(F)$, linking the obstacle velocity, $V$, only to the instantaneous applied load, $F$, quantitatively summarizes the combined action of the elementary self--assembling processes. 
In such adiabatic conditions, implying a time scale separation between the self--assembling process and the response of the obstacle, the $V(F)$ dependence could be probed equivalently by different protocols like, to cite the two most frequently used, the constant force load (e.g. clamped force set--up), where one directly observes the steady state velocity, and the harmonic load (the sample grows against an AFM cantilever or an optical trap), where the obstacle velocity can be followed as the load increases continuously up to stalling.

Abiabatic conditions cannot be in general guaranteed and indeed, careful investigations on a branched actin network growing against an AFM tip have shown that the recorded $V(F)$ relationship can be function of the load history \cite{Fletcher.2005}. 
The direct force--velocity relationship, $V(F)$, is in any event widely used as a characteristics of network dynamics to compare experimental measurements and modeling approaches for \textit{in--vitro} \cite{Smith2013} and \textit{in--vivo} systems \cite{Schreiber2010}.

To make progress on the rationalization of the conditions of validity of the widely used concept of force--velocity relationship, $V(F)$, we will restrict our considerations to a simple network where a bundle of parallel (proto)filaments (actin or tubulin) grows normally against a loaded obstacle. 
The general mechanism, linking work production and (de)polymerization kinetics of living bio--filaments, has been originally formulated theoretically by Hill for an incompressible bundle of $N_f$ parallel filaments pressing against a mobile obstacle \cite{Hill}. 
Successively, when the filaments of the bundle are treated as independent and equivalent and when it is assumed that the depolymerization rate is unaffected by the external load, the wall velocity $V^{MF}$ ($MF$ indicates the mean field character of this treatment) has been written as \cite{Borisy2008,vanDoorn2000}
\beq
V^{MF}(F;\rho_1,N_f)=d\left[U_0\exp\left(-\frac{Fd}{N_fk_BT}\right)-W_0\right]
\label{vmf_eq}
\eeq
where $U_0=k_{on}\rho_1$ and $W_0=k_{off}$ are the single filament bulk rate constants, related to bulk chemical rate constants $k_{on}$ and $k_{off}$, for single monomer polymerization and depolymerization steps, $\rho_1$ is the free monomer density, $d$ is the single filament increment of contour length per incorporated monomer and $F$ is the external force exerted on the wall. 
Supercritical conditions, where filament polymerization dominates over depolymerization, require $\hrho=\frac{\rho_1}{\rho_{1c}}=\frac{U_0}{W_0}>1$ where $\rho_{1c}=\frac{k_{off}}{k_{on}}$ is the critical value of the monomer density at which the bundle has no tendency to grow nor to shrink in absence of load.

Eq.\eqref{vmf_eq} predicts, for $F=0$, a growth velocity of the free bundle $V^{MF}=d\left(U_0-W_0\right)>0$  while the stalling force $F_s$, at which the velocity vanishes, is given by
\beq
F_s^{H}=N_f\frac{k_BT}{d}\ln\hrho.
\label{eq:Fstal}
\eeq
The notation $F_s^{H}$ reminds that this expression was originally established by Hill using thermodynamic arguments \cite{Hill1982}. 
Eq.\eqref{eq:Fstal} has been recently derived, in a special limit, by equilibrium Statistical Mechanics for a bundle of rigid filaments \cite{Paper2}. 
Indeed, it has been found that the statistical mechanics average of the wall position, taken over the equilibrium optical trap ensemble, multiplied by $\kappa_T$, converges exponentially fast for $\kappa_T \to 0$ to the Hill's prediction.

Experimental measurements of the force--velocity relationship for multi--filament bundles (tubulin or actin) \cite{Dogterom.1997,vanDoorn2000,Dogterom.07,DCBB.14} are not many, reflecting the difficulty to prepare \textit{in--vitro} the grafted bundle seed needed to follow its subsequent loaded growth. However, it is interesting to note the diversity in these few approaches. The growth of single grafted tubulin filaments, which are bundles of $13$ proto--filaments, was followed by imaging techniques \cite{Dogterom.1997,vanDoorn2000}. Regrouping $(F,V)$ data for different observation times and for different samples, a master force--velocity relation could be established. In another experiment using an acrosome bead complex of $N_f=8\div10$ F--actin filaments held in an optical trap device, the growth of a bundle was followed in time against an harmonic load \cite{Dogterom.07}. A rising signal finishing with a plateau was observed but the final stationary force was surprisingly much lower than the expected stalling force, Eq.\eqref{eq:Fstal}, its value being close to the stalling force predicted for a single filament. 
The analysis in this experiment considers many relaxation curves, but in many cases data had to be eliminated due to interferences during the relaxation process with the onset of escaping filaments. This happens because growing filaments undergo a large bending fluctuation which allows them to start growing freely along the obstacle. 
The transient behavior, which can be converted into a $V(F)$ law by the derivation in time of the wall position, was not exploited. 
Finally, in a recent study \cite{DCBB.14}, recording the rate of radial distance between two colloidal particles separated by a growing grafted actin bundle, the force--velocity relationship of actin bundles was established in constant load conditions. 

The outcome of the earliest experiment \cite{Dogterom.1997,vanDoorn2000}, confirmed by the more recent experimental work \cite{DCBB.14}, is that the velocity, and hence the power of transduction of multi filament bundles, is much lower than predicted by Eq.\eqref{vmf_eq}. 
The discrepancy highlights the non--independence of elementary chemical steps at the tip of different filaments in the bundle, with the effect of reducing the additivity of the action of each filament. 
The bundle model needs to be specified and the dependence between chemical events and wall position for a given longitudinal seeds disposition has to be quantitatively taken into account. 
This aspect is present in the multi--filament Brownian Ratchet (BR) models \cite{vanDoorn2000,DCBB.14,MO.1999,Tsekouras.2011} which generalize the single filament brownian ratchet model introduced by Peskin et al. \cite{Peskin}, for which one finds that the velocity vanishes for a load equal to Hill's expression, Eq.\eqref{eq:Fstal} \cite{vanDoorn2000}. For these bundle models, the important characteristics which distinguish the dynamical behavior of the bundle are the number of rigid living filaments, the longitudinal disposition of the seeds of the filaments and the wall diffusion coefficient $D$ which introduces a second characteristic time $\tau_D=d^2/D$ next to the chemical events time scale $\tau_{chem}=W_0^{-1}$. This fact suggests to introduce the parameter $\epsilon=\tau_D/\tau_{chem}$ to be able to discuss the condition of this second adiabatic separation (not to be confused with the one associated to the existence of $V(F)$). 
For both experiments having probed the $V(F)$ relationship, it was found that data could be interpreted successfully with the model of a staggered bundle (= staggered longitudinal seed disposition \cite{Paper2}) of rigid filaments in very fast wall diffusion conditions ($\epsilon=0$), a model we will denote as SRBR (Staggered Rigid Brownian Ratchet). 
On the contrary, for a similar model with an in registry bundle (unstaggered longitudinal seed disposition) \cite{Tsekouras.2011}, the predicted velocity was much too low with respect to the experimental data \cite{Dogterom.1997,vanDoorn2000,DCBB.14}.

In the stochastic dynamical models here considered, the force--velocity relationship depends parametrically, for a given seed arrangement, on the number of filaments, the reduced free monomer concentration and the time scale ratio $\epsilon$. In the case of constant load, the explicit form for the asymptotic force--velocity relationship, $V^{stat}(F;N_f,\hat{\rho}_1,\epsilon)$, for our models has been established by stochastic dynamics studies at finite $\epsilon$ \cite{Wang-Carlsson} and at $\epsilon=0$ \cite{vanDoorn2000,DCBB.14}. In the latter case a simplified algorithm, exploiting the time scales separation, has been used for the staggered bundle case. Indeed, at $\epsilon=0$, the wall position distribution at given filaments configuration, is found to be time--independent and equal to the equilibrium distribution of the wall position resulting from the 1D Brownian motion of a wall in the external load field, with the wall positions restricted to be greater than the position of the most advanced filament tip.

Interestingly, we add that for the SRBR model ($\epsilon=0$), successive theoretical developments \cite{MO.1999,vanDoorn2000,DCBB.14} have given, with a very good approximation \cite{DCBB.14}, two coupled closed expressions for the velocity $V^{stat}(F;N_f,\hat{\rho}_1)$ and the distribution of filament relative sizes (see Section \ref{ResSec}) $g(k;F,N_f,\hat{\rho}_1)$, for the stationary state.

In this work, we consider the stochastic staggered bundle model of rigid filaments in supercritical conditions and perform a series of dynamical runs for different load conditions. 
We first look at the constant force case, treating both the stationary state itself and the asymptotic transient evolution to reach it. 
We next envisage the bundle, in similar thermodynamic conditions, initially taken with very short filaments, subject to a harmonic load $-\kappa_T L$, where $L$ is the wall position and where $\kappa_T$ is trap strength (optical trap set--up). 
Mimicking the optical trap experiment \cite{Dogterom.07}, the bundle and the average wall position grow and reach stalling. 
We compare our computed longest relaxation time with a theoretical approximate expression derived along the lines of the D\'emoulin et al. theory.
We derive and compare the force--velocity relationship extracted from this optical trap relaxation with the one obtained in stationary conditions. As expected, we found that the two coincide in adiabatic conditions, i.e. when the characteristic time of the optical trap relaxation is much larger then the characteristic time of the relaxation in the constant force case.

Our algorithms follow the same lines of those used in previous studies. 
However, in our study we deal with an optical trap load, while most studies (with an exception restricted to the $\epsilon=0$ case \cite{Carlsson}) assume a constant load.
Moreover, while algorithms for finite $\epsilon$ or $\epsilon=0$ are usually just assumed, we establish an explicit link showing how the $\epsilon=0$ model is derived from the general finite $\epsilon$ case.

In Section \ref{sec2} we present the general Fokker--Planck model for a bundle of rigid filaments with an arbitrary seed disposition, facing either a constant or a harmonic load and we derive the explicit wall algorithm (EWA) giving the sampling rules to generate stochastic trajectories for any finite $\epsilon$ case. 
We then use a perturbation expansion to derive the $\epsilon=0$ model, still for the constant load or the harmonic load, and we derive the simplified implicit wall algorithm (IWA) giving the sampling rules in the $\epsilon=0$ case. 
Section \ref{ResSec} reports and discusses our results for constant force and optical trap loads for the same bundle system generally using the $\epsilon=0$ approach, since the wall diffusion takes place very quickly with respect to the mean time between (de)polymerization events. However, we also verify that the simplified algorithm is robust, since we find identical results in a reasonable range of $\epsilon$ non--zero values.
Section \ref{concl_sec} concludes with a summary of the main results and with some perspectives.

\section{Model and Implementation}\label{sec2}

We consider a bundle of $N_f\geqslant1$ living filaments, grafted normally (say along the $x$ axis) to a fixed planar substrate wall (along $y$ and $z$ directions). 
The filaments are modeled as discrete rigid linear chains with monomer size $d$ and length related to the number of attached monomers, $i\geq2$, as $L_{ci}=(i-1)d$.
Let $h_n$ be the location along $x$ axis of the seed (first monomer) of the filament $n$ close to the grafting plane ($-d/2<h_n<d/2$).
For a bundle of many filaments, two seed dispositions are usually considered: \textit{in--registry} (or \textit{unstaggered}), where $h_n=0$, $n=1,N_f$, and \textit{homogeneous} (or \textit{staggered}), where seeds are regularly spaced as
\beq
h_n=\left[\frac{n-0.5}{N_f}-0.5\right]d\ \ \ n=1,N_f.
\eeq

A moving obstacle, a hard wall located at distance $L$ from the parallel substrate wall, is loaded with a compressional external force $F$ bringing it into contact with the living filaments.
We will consider two types of load, the constant force $F$, and the optical trap setting, with $F=-\kappa_TL$, where $\kappa_T$ is the trap stiffness and $L$ the distance between the walls.

The bundle force for rigid filaments is impulsive. 
Its effect is taken into account by imposing a confining boundary to the wall motion at the tip location of the longest filament.

Filaments either grow by a single monomer polymerization step with bulk rate $U_0$, proportional to the free monomer density $\rho_1$, or shrink by a single monomer depolymerization step with bulk rate $W_0$.
The ratio $U_0/W_0=\hrho$ is the free monomer density divided by its critical value, i.e. the value at which the two bulk rates are equal.
We will be interested to supercritical conditions only ($\hrho>1$), where the filaments tend to grow against the loaded wall.
When a filament tip gets closer than $d$ to the wall, the polymerization rate becomes zero, while the depolymerization one is assumed to remain unchanged. 

The dynamics of the bundle of growing filaments against the loaded mobile wall presents two main time scales: the chemical one, $\tau_{chem}=1/W_0\sim1/U_0$, and the diffusive one, related to the diffusive motion of the wall and estimated by $\tau_D=d^2/D$ where $D$ is the wall diffusion coefficient. The ratio of time scales, $\epsilon=\tau_D/\tau_{chem}$, in typical \textit{in--vitro} experiments is $\epsilon\ll1$, but might sometimes go close to 1 for a very large colloidal particle in a crowded environment.

In the following, we establish a Fokker--Planck equation to describe the dynamics of an arbitrary bundle of independent rigid filaments subjected to a constant or harmonic load, for arbitrary value of $\epsilon$.

\subsection{General Fokker--Plank equation for a bundle of rigid filaments against a constant or harmonic load}

We describe the time evolution of $N_f$ filaments against a load in terms of the filament sizes and the wall position, $\{j_1,\dots,j_{N_f},L\}$. 
The wall position must always lie beyond the tip of any filament -- and so beyond the tip of the most advanced one, $n^\ast$ with size $j_{n^\ast}$. Defining $X_n(j_n)$, the position of the tip of filament $n$ and $X^\ast$ that of the most advanced one,
\bea
X_n(j_n)&=&(j_n-1)d+h_n\label{Xn}\\
X^\ast\equiv X^\ast(j_1,\dots,j_{N_f})&=&\max_{n=1,N_f}\{X_n(j_n)\}=X_{n^\ast}(j_{n^\ast}),\label{Xstar}\\
L&\geqslant&X^\ast.
\eea
We assume that the joint probability distribution function $P_{j_1,\dots,j_{N_f}}(L,t)$ satisfies a Fokker--Planck equation in time mixing a continuous process in space for the wall position with a discrete process for filament sizes. For the model described above, we have
\bea
&&\frac{\partial P_{j_1,\dots,j_{N_f}}(L,t)}{\partial t} +\frac{\partial}{\partial L} J_{j_1,\dots,j_{N_f}}(L,t)=\nonumber\\
 &&U_0\left[\sum_{n=1}^{N_f}(1-\delta_{2,j_n})\Theta\left(L-X_n(j_n)\right)P_{j_1,\dots,j_n-1,\dots,j_{N_f}}(L,t)-\sum_{n=1}^{N_f}\Theta\left(L-X_n(j_n+1)\right)P_{j_1,\dots,j_n,\dots,j_{N_f}}(L,t)\right]\nonumber\\
 &+&W_0\left[\sum_{n=1}^{N_f}P_{j_1,\dots,j_n+1,\dots,j_{N_f}}(L,t)-\sum_{n=1}^{N_f}(1-\delta_{2,j_n})P_{j_1,\dots,j_n,\dots,j_{N_f}}(L,t)\right]
\label{FPeq}
\eea
where $\Theta(x)$ is the Heaviside step function and the probability current density is
\bea
J_{j_1,\dots,j_{N_f}}(L,t)&=&-D\left[\frac{\partial P_{j_1,\dots,j_{N_f}}(L,t)}{\partial L}- \frac{F(L)}{k_BT} \;\;P_{j_1,\dots,j_{N_f}}(L,t)\right]
\label{eq:pcdF}
\eea
In Eq.\eqref{eq:pcdF} the compressive force can be either a constant $F<0$ or an elastic force $F(L)=-\kappa_TL$ modeling the optical trap.
The right--hand side of Eq.\eqref{FPeq} represents the sink and source terms affecting the dynamics due to polymerization and depolymerization events.
Their explicit expression indicates that, in one step at fixed $L$, transitions are only possible between adjacent microscopic states, where $(N_f-1)$ filaments have the same size while the size of the remaining filament differs by $\pm1$ unit, taking into account the restriction $L\geqslant X^\ast$, and that the filament size cannot be smaller than two.

The general normalization condition for the distribution $P_{j_1,\dots,j_{N_f}}(L,t)$ is 
\beq
\sum_{j_1=2}^{\infty}\cdots\sum_{j_{N_f}=2}^{\infty}\int_{X^\ast}^\infty dL\; P_{j_1,\dots,j_{N_f}}(L,t)=1
\label{eq:norm}
\eeq
while the boundary conditions on the probabilities are  
\bea
P_{j_1,\dots,j_{N_f}}(L,t)\big|_{L<X^\ast}&=&0 \qquad P_{j_1,\dots,j_{N_f}}(L,t)\big|_{L=\infty}=0\label{bcP}\\
J_{j_1,\dots,j_{N_f}}(L,t)\big|_{L=X^\ast}&=&0 \qquad J_{j_1,\dots,j_{N_f}}(L,t)\big|_{L=\infty}=0\label{bcJ}
\eea

To simplify the treatment of the continuous--discrete structure of Eq.\eqref{FPeq}, we discretize, following reference \cite{Wang2003}, the wall position with a grid step $\delta=d/M$, with $M$, integer, $\gg1$. 
We then substitute to the wall position $L$ the discrete variable
\beq
k=\integer\left[\frac{L}\delta\right]\equiv\integer\left[l\right].
\label{Ladim}
\eeq
In this way Eq.\eqref{FPeq} will become a finite difference equation in $k$ representing a discrete Markov chain in continuous time
\beq
\frac{d\mathbf{\mathcal{P}}}{dt}=\mathbf{\mathcal{P}}\mathbf{Q}
\label{FPdiscr}
\eeq
with $\mathcal{P}(t)=\{\mathcal{P}_{j_1,\dots,j_{N_f},k}(t)\}_{j_n\in\left[2,\infty\right)\ n=1,N_f,k\in\left[\integer\left[(d+h_{N_f})/\delta\right],\infty\right)}$ a vector field and $\mathbf{Q}$ the generator matrix of the Markov chain. 
The elements of the matrix $\mathbf{Q}$ 
contain the (de)polymerization rates for the filaments, 
\bea
U_{j_n}(L)&=&U_0\Theta\left(L-X_n(j_n+1)\right)\label{polrate}\\
W_{j_n}(L)&=&W_0\label{depolrate}
\eea
and the forward/backward jump rates for the wall; the expressions of these matrix elements are given in Appendix \ref{appA}.

To circumvent the difficulty of solving analytically Eq.\eqref{FPdiscr}, one can produce a number of realizations of the discrete Markov chain using any appropriate algorithm, in our case the Gillespie algorithm \cite{Gillespie,GG.07}: given an initial condition at time $t_0$, the state of the system is estimated in terms of the set of random variables $\{j_1,\dots,j_{N_f},k\}$ at time $t$ producing statistically correct trajectories, from which the probability distribution function $\mathcal{P}_{j_1,\dots,j_{N_f},k}(t)$ can be inferred by histograms. 
Starting from the initial state, the system is allowed to evolve by random steps involving only one reaction per time: one filament depolymerization or polymerization, or the wall forward or backward jump. 
Denoting by $i_0$ the current state of the system, the reachable states $i_m$ are those differing from $i_0$ for only one variable by $\pm1$, namely $\{j_1,\dots,j_n\pm1,\dots,j_{N_f},k\}$ or $\{j_1,\dots,j_n,\dots,j_{N_f},k\pm1\}$. 
It is straightforward to see that the number of these possible final states is $2N_f+2$.
The transitions $i_0\to i_m$, $m\in\left[1,2N_f+2\right]$, are described in Eq.\eqref{FPdiscr} by the generator matrix elements $Q_{i_mi_0}$, the rates of going from $i_0$ to $i_m$. 
The corresponding diagonal element is $Q_{i_0i_0}=-\sum_{i_m\neq i_0}Q_{i_mi_0}$ \cite{Norris}.
The evolution of the system is determined by two random variables: the time to the next reaction, $\tau$, and the final state $i_m$, or equivalently the index of the  reaction, $m\in\left[1,2N_f+2\right]$. 
From general Markov chain theory, $\tau$ is known to be an exponentially distributed random variable: given the current state $i_0$, the parameter of the exponential distribution is given by $-Q_{i_0i_0}$. 
Instead, the probability for the jump $m$ linking states $i_0$ and $i_m$ to take place is given by the ratio between $Q_{i_0i_m}$ and $|Q_{i_0i_0}|$ \cite{Norris}.
The main loop of the algorithm follows this scheme:
\begin{enumerate}
\item[0.] The initial state $i_0$ is specified in terms of the state vector $\{j_1,\dots,j_{N_f},k\}$. We take for the initial value of $k$ a small fixed value and, for the filament, compatible initial sizes; 
\item[1.] The matrix elements $Q_{i_0i_m}$ are calculated for any state $i_m$ reachable from $i_0$; 
\item[2.] The time to the next move is determined using the so--called direct method, which follows from the standard inversion method of the Monte Carlo theory \cite{GillespieBook}: a random number $r_1\in\left[0,1\right]$ is generated from the uniform distribution and the time $\tau$ is taken as 
\beq
\tau=\frac{1}{|Q_{i_0i_0}|}\ln\frac1{r_1};
\eeq
\item[3.] The index of the next move is determined using the same method: a second random number $r_2\in\left[0,1\right]$ is generated and the index $m$ is taken as the smallest integer satisfying 
\beq
\sum_{n=1}^{m-1}\frac{Q_{i_0i_n}}{|Q_{i_0i_0}|}<r_2\leqslant \sum_{n=1}^{m}\frac{Q_{i_0i_n}}{|Q_{i_0i_0}|};
\eeq
\item[4.] The sampled move is taken by updating the state vector $i_0\to i_m$ and the time is incremented by $\tau$;
\item[5.] Go back to 1, until a maximum time $t_{max}$ is reached;
\item[6.] End the simulation.
\end{enumerate}
The state vector $\{j_1,\dots,j_{N_f},k\}$ is stored for the calculation of histograms and averages.

This algorithm \cite{Gillespie,Wang2003}, solving the Fokker--Planck Eqs.\mref{FPeq,eq:pcdF}, works for any seed disposition (staggered and unstaggered), for any finite value of the dimensionless parameter $\epsilon\equiv\frac{d^2W_0}{D}$ and both for the two cases of constant force and optical trap load.
We will call it the explicit wall algorithm (EWA).

In the next subsection we treat the specific, important, reference case of loaded bundles of rigid filaments in the limit $\epsilon\to0$. 
In this limit the wall re-equilibrates instantaneously after any change of the position of the most advanced tip of the bundle. 
The interest of this limit is justified since in $in-vitro$ experiment with actin bundles / colloidal particles (e.g. the optical trap experiment \cite{Dogterom.07}) the typical value of the ratio of time scales is $\epsilon\ll1$. 
We will see that the dynamics of the bundle simplifies for two reasons: the elimination of the fast motion of the wall permits to go to longer times and the dimensionality of the problem is reduced. 
The new algorithm, called implicit wall algorithm (IWA), becomes then decidedly more efficient.

\subsection{Treatment of the Fokker--Planck equation in the $\epsilon=0$ limit \cite{kozyreff}}
Given the separation of time scales between the chemical events and the wall diffusion, it is convenient to rewrite Eq.\eqref{FPeq} in terms of dimensionless variables in order to put in evidence  the ratio $\epsilon=\frac{\tau_D}{\tau_{chem}}=\frac{W_0d^2}D$.
Defining $\tilde{t}=W_0t$, $x=\frac{L}d$ and $f=\frac{Fd}{k_BT}$, multiplying Eq.\eqref{FPeq} by $\frac{d^2}D$ and redefining the probability distribution functions, we get:
\bea
&&\epsilon\frac{\partial \widetilde{P}_{j_1,\dots,j_{N_f}}(x,\tilde{t})}{\partial \tilde{t}} +
\frac{\partial}{\partial x}\widetilde{J}_{j_1,\dots,j_{N_f}}(x,\tilde{t})=\nonumber\\
 &&\epsilon\Bigg\{\hrho\left[\sum_{n=1}^{N_f}(1-\delta_{2,j_n})\Theta\left(x-X_n(j_n)/d\right)\widetilde{P}_{j_1,\dots,j_n-1,\dots,j_{N_f}}(x,\tilde{t})-\sum_{n=1}^{N_f}\Theta\left(x-X_n(j_n+1)/d\right)\widetilde{P}_{j_1,\dots,j_n,\dots,j_{N_f}}(x,\tilde{t})\right]\nonumber\\
 &+&\sum_{n=1}^{N_f}\widetilde{P}_{j_1,\dots,j_n+1,\dots,j_{N_f}}(x,\tilde{t})-\sum_{n=1}^{N_f}(1-\delta_{2,j_n})\widetilde{P}_{j_1,\dots,j_n,\dots,j_{N_f}}(x,\tilde{t})\Bigg\}
\label{FPadim}
\eea
with 
\beq
\widetilde{J}_{j_1,\dots,j_{N_f}}(x,\tilde{t})=-\frac{\partial}{\partial x}\widetilde{P}_{j_1,\dots,j_{N_f}}(x,\tilde{t}) -f(x)\widetilde{P}_{j_1,\dots,j_{N_f}}(x,\tilde{t})
\eeq
the probability current density in the reduced units.
In the $\epsilon\to0$ limit, it is legitimate to replace Eq.\eqref{FPadim} by its simpler $\epsilon$ zero--th order approximation:
\beq
\frac{\partial^2\widetilde{P}^{(0)}_{j_1,\dots,j_{N_f}}(x,\tilde{t})}{\partial x^2} -\frac{\partial}{\partial x}\left[-f(x)\widetilde{P}^{(0)}_{j_1,\dots,j_{N_f}}(x,\tilde{t})\right]=0.
\label{FPzero}
\eeq
By integrating in $dx$ form $X^*$ to $\infty$ and using the boundary conditions for the probability, one gets $\frac{\partial\widetilde{P}^{(0)}}{\partial x}=-f(x)\widetilde{P}^{(0)}$ so that the general solution is: 
\beq
\widetilde{P}^{(0)}_{j_1,\dots,j_{N_f}}(x,\tilde{t})=a(j_1,\dots,j_{N_f},\tilde{t})\exp\left(-\int_x^\infty dx f(x)\right)
\label{gensol}
\eeq
On the other side, it is always possible to write the joint probability as the product of the marginal distribution for the subset $\{j_1,\dots,j_{N_f}\}$ times the conditional probability distribution for $x$:
\beq 
\widetilde{P}^{(0)}_{j_1,\dots,j_{N_f}}(x,\tilde{t})=\widetilde{P}_0(j_1,\dots,j_{N_f},\tilde{t})\widetilde{P}_0(x\;|\;j_1,\dots,j_{N_f},\tilde{t}) 
\eeq
Therefore, given the general solution \eqref{gensol}, we can write it as 
\beq
\widetilde{P}^{(0)}_{j_1,\dots,j_{N_f}}(x,\tilde{t})=\widetilde{P}_0(j_1,\dots,j_{N_f},\tilde{t})\widetilde{P}_{EQ}(x\;|\;j_1,\dots,j_{N_f}) \label{Peq0}
\eeq
as the $x$ dependence, Eq.\eqref{gensol}, is explicit and time--independent.
The wall distribution $\widetilde{P}_{EQ}(x\;|\;j_1,\dots,j_{N_f})$ is an explicit, time independent, normalized, distribution for the wall position conditional to the set of filaments sizes.
Explicit expression for the two normalized cases of constant load and optical trap are:
\beq
\widetilde{P}_{EQ}(x\;|\;j_1,\dots,j_{N_f})=\begin{dcases*}\frac{f\exp(-fx)}{\exp(-fX^\ast/d)}\qquad &\mbox{constant load}\\
\sqrt{\frac{2\tilde{\kappa}_T}\pi}\frac{\exp\left(-\frac12\tilde{\kappa}_Tx^2\right)}{\erfc\left(\sqrt{\frac12\tilde{\kappa}_T}X^\ast/d\right)}\qquad &\mbox{optical trap}
\end{dcases*}
\label{Peq}
\eeq
with $\tilde{\kappa}_T=\frac{\kappa_Td^2}{k_BT}$.
From Eq.\eqref{Peq} we get the average wall position conditional to the bundle sizes $\{j_1,\dots,j_{N_f}\}$ as:
\beq
E(x\;|\;j_1,\dots,j_{N_f})=\int_{X^*}^\infty x\widetilde{P}_{EQ}(x\;|\;j_1,\dots,j_{N_f})dx=\begin{dcases*}\frac{X^\ast}d+\frac1f\qquad &\mbox{constant load}\\
\sqrt{\frac2{\tilde{\kappa}_T\pi}}\frac{\exp\left[-\frac12\tilde{\kappa}_T\left(X^\ast/d\right)^2\right]}{\erfc\left(\sqrt{\frac12\tilde{\kappa}_T}X^\ast/d\right)}\qquad &\mbox{optical trap}
\end{dcases*}
\label{Lav}
\eeq

Note that the full distribution at $\epsilon=0$, given by Eq.\eqref{Peq0}, is still a time--dependent function, since filament sizes change by single monomer polymerization/depolymerization events; the infinite separation of the time scales ($\epsilon=0$) implies that after any chemical event, the wall immediately re--equilibrates according to the time--independent distribution, Eq.\eqref{Peq}, given the new set of filament sizes.
To get the full distribution, we write $\widetilde{P}_{j_1,\dots,j_{N_f}}(x,\tilde{t})$ as an asymptotic expansion in terms of the small parameter $\epsilon$:
\beq
\widetilde{P}_{j_1,\dots,j_{N_f}}(x,\tilde{t})=\widetilde{P}^{(0)}_{j_1,\dots,j_{N_f}}(x,\tilde{t})+\epsilon\widetilde{P}^{(1)}_{j_1,\dots,j_{N_f}}(x,\tilde{t})+\dots
\eeq
where $\widetilde{P}^{(0)}_{j_1,\dots,j_{N_f}}(x,\tilde{t})$ is given by Eq.\eqref{Peq0}.
If we substitute this expansion, truncated to the first order, into Eq.\eqref{FPadim}, to the order $\epsilon$ we find the following equation:
\bea
&&\frac{\partial \widetilde{P}^{(0)}_{j_1,\dots,j_{N_f}}(x,\tilde{t})}{\partial \tilde{t}} +\frac{\partial}{\partial x}\;\widetilde{J}^{(1)}_{j_1,\dots,j_{N_f}}(x,\tilde{t})=\nonumber\\
 &&\hrho\left[\sum_{n=1}^{N_f}(1-\delta_{2,j_n})\Theta\left(x-X_n(j_n)/d\right)\widetilde{P}^{(0)}_{j_1,\dots,j_n-1,\dots,j_{N_f}}(x,\tilde{t})-\sum_{n=1}^{N_f}\Theta\left(x-X_n(j_n+1)/d\right)\widetilde{P}^{(0)}_{j_1,\dots,j_n,\dots,j_{N_f}}(x,\tilde{t})\right]\nonumber\\
 &+&\sum_{n=1}^{N_f}\widetilde{P}^{(0)}_{j_1,\dots,j_n+1,\dots,j_{N_f}}(x,\tilde{t})-\sum_{n=1}^{N_f}(1-\delta_{2,j_n})\widetilde{P}^{(0)}_{j_1,\dots,j_n,\dots,j_{N_f}}(x,\tilde{t})
\label{FPone}
\eea
Integrating both sides of this equation from $x=X^\ast/d$ to $\infty$, applying the boundary conditions Eq.\eqref{bcJ} on $\widetilde{J}^{(1)}$ and the normalization of $\widetilde{P}_{EQ}$ and using Eq.\eqref{Peq0}, we get:
\bea
&&\frac{\partial \widetilde{P}_0(j_1,\dots,j_{N_f},\tilde{t})}{\partial \tilde{t}}=\nonumber\\
 &&\hrho\sum_{n=1}^{N_f}(1-\delta_{2,j_n})\int_{X^\ast/d}^\infty dx\;\Theta\left(x-X_n(j_n)/d\right)\widetilde{P}_{EQ}(x\;|\;j_1,\dots,j_n-1,\dots,j_{N_f})\widetilde{P}_0(j_1,\dots,j_n-1,\dots,j_{N_f},\tilde{t})\nonumber\\
 &-&\hrho\sum_{n=1}^{N_f}\int_{X^\ast/d}^\infty dx\;\Theta\left(x-X_n(j_n+1)/d\right)\widetilde{P}_{EQ}(x\;|\;j_1,\dots,j_n,\dots,j_{N_f})\widetilde{P}_0(j_1,\dots,j_n,\dots,j_{N_f},\tilde{t})\nonumber\\
 &+&\sum_{n=1}^{N_f}\widetilde{P}_0(j_1,\dots,j_n+1,\dots,j_{N_f},\tilde{t})-\sum_{n=1}^{N_f}(1-\delta_{2,j_n})\widetilde{P}_0(j_1,\dots,j_n,\dots,j_{N_f},\tilde{t})
\label{FPrho_one}
\eea
where we couldn't use the normalization condition for the terms where we have left the integration explicitly written.
This equation describes a discrete process for the filament sizes in continuous time, which can be rewritten in a vectorial form, similar to Eq.\eqref{FPdiscr}:
\beq
\frac{dP_0}{dt}=P_0\mathbf{Q}^{(0)}
\label{CTMC0}
\eeq
with $\mathbf{Q}^{(0)}$ generator matrix of the process, whose elements are given in Appendix \ref{appB}. 

The numerical solution of the Markov chain equation described by Eq.\eqref{CTMC0} follows exactly the same scheme described above for the general Fokker--Planck equation for $\epsilon>0$. 

As already mentioned, in this case the algorithm is more efficient since it spans longer times (we have integrated out the fast variable) and it has to treat a reduced number of variables.

The solution of Eq.\eqref{CTMC0} and the conditional probability for the wall position Eq.\eqref{Peq}, give the necessary information needed to compute all time-dependent ensemble averages, as e.g. $\<L\>_t$. Similar model and procedures have been used: i. for constant load option and in-registry \cite{Tsekouras.2011} or staggered \cite{MO.1999,vanDoorn2000,DCBB.14} bundles; ii. for optical trap only for staggered bundles \cite{Carlsson}.

\section{Simulations and Results}\label{ResSec}

\subsection{Units, parameters and stochastic runs}

In our simulations, length, time and energy units are taken as $d$, $W_0^{-1}$, and $k_BT$ respectively. All quantities will be mentioned in reduced units based on the above three fundamental units.
For actin $d=2.7~nm$; experimental information for $W_0$ gives $ W_0=1.4~s^{-1}$;
and, at room temperature $k_BT=4.14\times10^{-21}~J$. We choose to perform our studies on a bundle of $N_f=32$ rigid filaments with a staggered disposition of seeds at a reduced density $\hrho=\frac{U_0}{W_0}=2.5$. 
With reference to a wall constituted by a bead of micron size in water opposing the actin bundles \cite{Dogterom.07,DCBB.14}, experimental information gives for the adimensional parameter introduced in the previous section, the value $\epsilon=5.5\times10^{-5}$.
Given the small value of $\epsilon$, we performed the major part of our simulations in the $\epsilon =0$ limit with the IWA algorithm. However, we have considered interesting to compare the results of the IWA algorithm with those of the EWA corresponding to a finite but small value of $\epsilon$.
With the very small experimental value of $\epsilon$, EWA would be highly inefficient, since the computer time would be essentially spent to study the wall diffusion next to a bundle with quasi-fixed filament sizes. 
Since for the load-velocity relationship we need to sample both wall and filament sizes, we decided to adopt a value of epsilon thousand times bigger, $\epsilon=5\times10^{-2}$. This value, in fact, still permits to give a sufficient representation of the wall dynamics. Our EWA approach requires to discretize the space variable $L$ with elementary steps $\delta=d/M$. For $M$, we have adopted $M=100$. 
To compute the solution of our Fokker--Planck equation, both for $\epsilon=0$ (IWA) or $\epsilon>0$ (EWA), we need to fix initial configurations. Our choice for EWA has been to fix the wall location $L_0$ (i.e. $k_0=\integer(L_0/\delta)$) and to sample the initial filament sizes for each trajectory of the stochastic dynamics according to the filament size equilibrium probability $P^{eq}(j_1,j_2,....j_n,....j_{N_f};L_0)$, conditional to the chosen wall location \cite{Paper2}. For initiating IWA runs, the initial filament sizes must be arbitrarily chosen and the initial wall location then follows from its conditional distribution.

\subsection{Observables of interest}
\begin{enumerate}
\item[1)] Wall position
\end{enumerate}
The wall position $L$ is the quantity directly followed in time in real experiments and corresponds to the expected value of the random variable $\hat{L}$ over the solution of the FP equation, $\<\hat{L}\>_t$. 
The calculation of this quantity is direct in the EWA case, while it has to be determined in the IWA case through the instantaneous size distribution of $j_n$, $n=1,32$, implying $\hat{L}$ values ahead of the tip of the most advanced filament at $X^*$, given by Eq.\eqref{Xstar}, using Eq.\eqref{Lav}.

\begin{enumerate}
\item[2)] Relative size (in number of monomers) of filaments with respect to the leading one.
\end{enumerate}
In terms of the tip positions $X_n$ and $X^*$ defined by Eqs.\mref{Xn,Xstar}, let us define the relative subset index $m=1,N_f-1$ given by
\beq
m(n)=\mod\left(\frac{X^*-X_n}{d/N_f},N_f\right)=\mod\left((j_{n^*}-j_n)N_f+n^*-n,N_f\right)\qquad n=1,N_f ; \;n\neq n^{*}
\eeq
This index represents in successive order the filament of order $n$ nearest neighbor of $n^*$, second neighbor of $n^*$, etc. Therefore it gives an intrinsic order to the vector representing the relative size of each filament. Note that the dividend in the function $\mod$ in the given condition is always positive. 
Then we can define, for each filament $n$, the quantity 
\beq
k_m=\integer\left[\frac{X^{*}-X_{n(m)}}{d}\right] =\integer\left[j_{n^*}-j_{n(m)}+\frac{n^*-n(m)}{N_f}\right] \qquad m=1,N_f-1  \label{eq:kn}
\eeq
Each component of this vector represents in discrete units of monomer size $d$ the relative distance from the most advanced tip of the first, second, etc. neighboring index.

This vector of relative sizes is interesting because its time-dependent probability distribution reaches a stationary value in the case of the wall subjected to a constant load.
\begin{enumerate}
\item[3)] Density of relative size of $N_f-1$ filaments with respect to the leading one
\end{enumerate}
This quantity is defined by the microscopic observable
\bea
\hat{g}(k)&=\frac{1}{(N_f-1)}\sum_{m=1}^{N_f-1} \delta_{k,k_m}\ \ \label{eq:gk}
\eea

At time $t$, the microscopic distribution will be $g(k,t)=\<\hat g(k)\>_t$. Specifically, we will characterize the internal structure of the bundle either by $g(0,t)$, the average probability densitythat the tip lies at a distance smaller than $d$ from the tip of the most advanced filament, or the average relative size $\<k\>_t=\sum_{k=0}^\infty kg(k,t)$. We will denote by $g(k)$ and $k_{av}$ the time--asymptotic values of these quantities for constant load force dynamics.

\subsection{Constant force load}

We have computed the relaxation towards the stationary state for a homogeneous bundle of $N_f=32$ rigid living filaments at $\hat{\rho}_1=2.5$ pressing against a constant load $F$. 
We have chosen various values of $F$ in the range $0.05<F/F_s< 1.25$ with $F_s$ the stalling force, Eq.\eqref{eq:Fstal}.
For each load value, we have generally used the IWA algorithm to produce $10^4$ independent trajectories, starting at time $0$ with all filament sizes set to the same value $(j_n(0)=500, n=1,32)$. We have chosen this value to avoid to fall at later times at the lower boundary $j_n=2$. That could happen when $F>F_s$ with negative average velocities. 
In two cases, starting with $L_0=5d$, we have used the EWA algorithm, averaging over $10^3$ independent trajectories. 
To determine the microscopic relaxation time of the bundle, we have fitted the asymptotic time evolution of the average wall position as $\<L\>_t=C+V^{stat} t + C'\exp{(-t/\tau_{micro})}$. 
To get the diffusion coefficient of the bundle $\Gamma$, we have also fitted the asymptotic behavior of the mean square elongation $\sigma^2(t)=\langle \hat{L}^2\rangle_t-\langle\hat{L}\rangle_t^2\isEquivTo{t\to \infty}2\Gamma t$ \cite{ranjith}.

\begin{figure}
\centering
\includegraphics[width=0.6\textwidth]{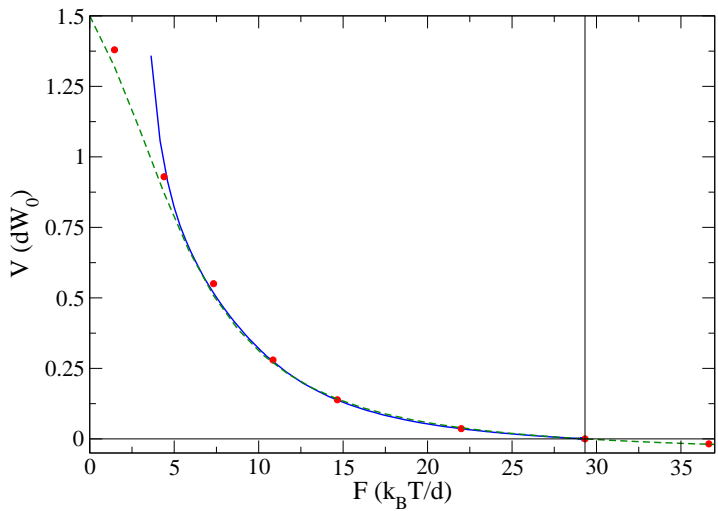}
\caption{Force--velocity relationship for a homogeneous bundle of $N_f=32$ rigid filaments at $\hat{\rho}_1=2.5$ ($\epsilon=0$). The $V^{stat}$ stationary velocity data points (red filled circles) are obtained as the asymptotic slope of $\<\hat{L}\>_t$ for constant force runs at each shown load value. Error bars are less than symbol sizes. The dashed green line is the D{\'e}moulin et al. theoretical estimate of $V^{stat}(F)$ based on Eqs.\mref{vf_dem,g_k}. The blue continuous curve is the force--velocity relationship obtained by the optical trap relaxation at $\kappa_T=0.4511$ (see text). Stalling is indicated by the vertical line at $F=29.32$.}
\label{rigid_vF}
\end{figure}

In Figure \ref{rigid_vF}, we report $V^{stat}(F)$ together with the D{\'e}moulin et al. prediction (Eqs.\mref{vf_dem,g_k}) for a similar staggered bundle of rigid filaments at $\epsilon=0$ in the same conditions \cite{DCBB.14}. 
This comparison shows that the theoretical prediction of $V^{stat}(F)$ represents quite accurately (the difference never exceeding $2\%$) the exact results obtained between zero load and stalling conditions.

\begin{figure}[ht]
\begin{center}
\includegraphics[width=0.6\textwidth]{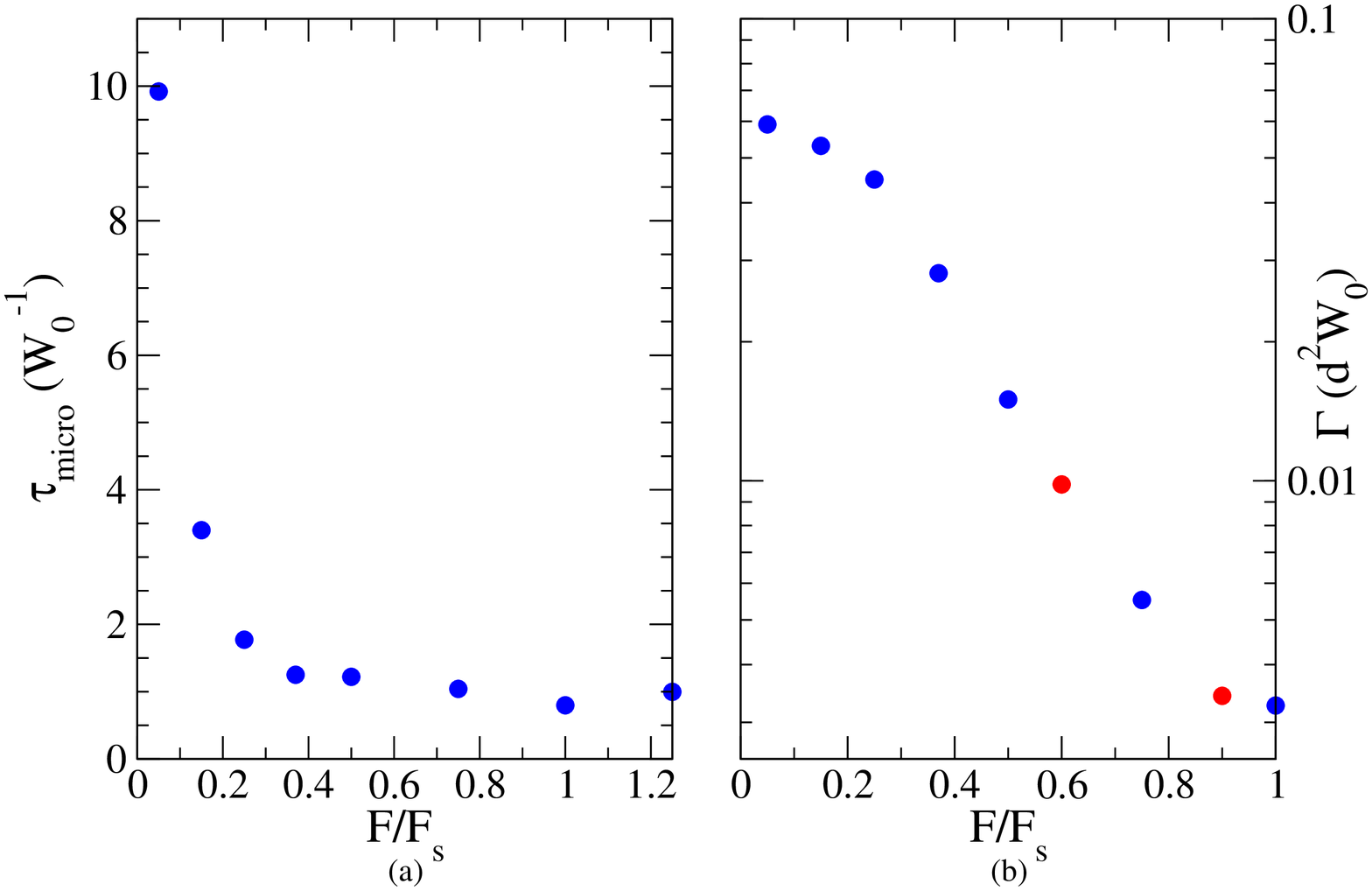}
\caption{(a) Load dependence of the relaxation time $\tau_{micro}$ for a homogeneous bundle of $N_f=32$ rigid filaments countering a constant load $F$ at $\hat{\rho}_1=2.5$. (b) Diffusion coefficient $\Gamma$ of the bundle. Blue symbols (IWA) and red symbols (EWA) refer to the stationary part of the constant load stochastic dynamics experiment mentioned in (a).} 
\label{fig:SM_tmicro}
\end{center}
\end{figure}

In Figure \ref{fig:SM_tmicro}, we collect transient times $\tau_{micro}$ and the diffusion coefficient of the bundle, $\Gamma$. Note the consistency within $\Gamma$ values obtained from IWA or EWA runs. $\tau_{micro}$ results of the order of $W_0^{-1}$ except at small loads where it diverges: in the discussion section we will come back to this important point.

\begin{figure}
\centering
\includegraphics[width=0.6\textwidth]{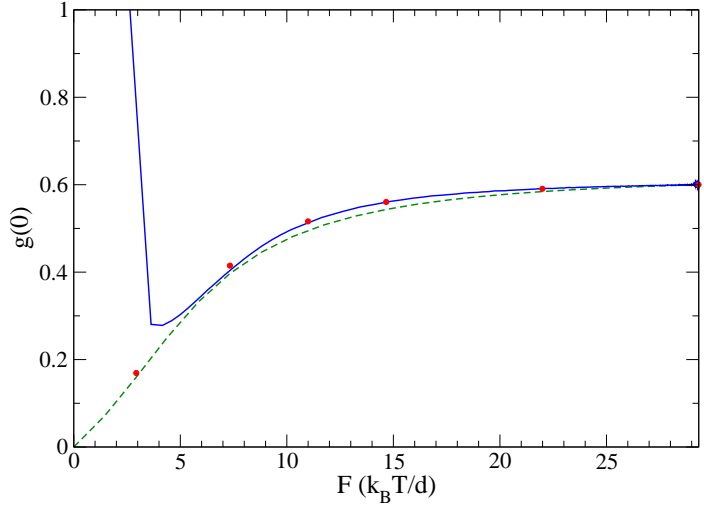}
\caption{Values at $k=0$ of the relative size distribution, g(0), as a function of the external load with $N_f=32$, $\hrho=2.5$ and $\epsilon=0$. The red filled circles are obtained in the stationary regime of constant load runs at the shown values of $F$. Error bars are less than symbol sizes. The dashed green line is the D{\'e}moulin et al. estimate of $g(0;F)$ based on Eq.\eqref{g_k}. The blue continuous curve is obtained for the optical trap  by eliminating from $g(0,t)=\<\hat{g}(0)\>_t$ and $\<F\>_t=\kappa_T \<L\>_t$ at $\kappa_T=0.4511$, the time parameter $t$.}
\label{g0data}
\end{figure}
 
\begin{figure}
\centering
\includegraphics[width=0.6\textwidth]{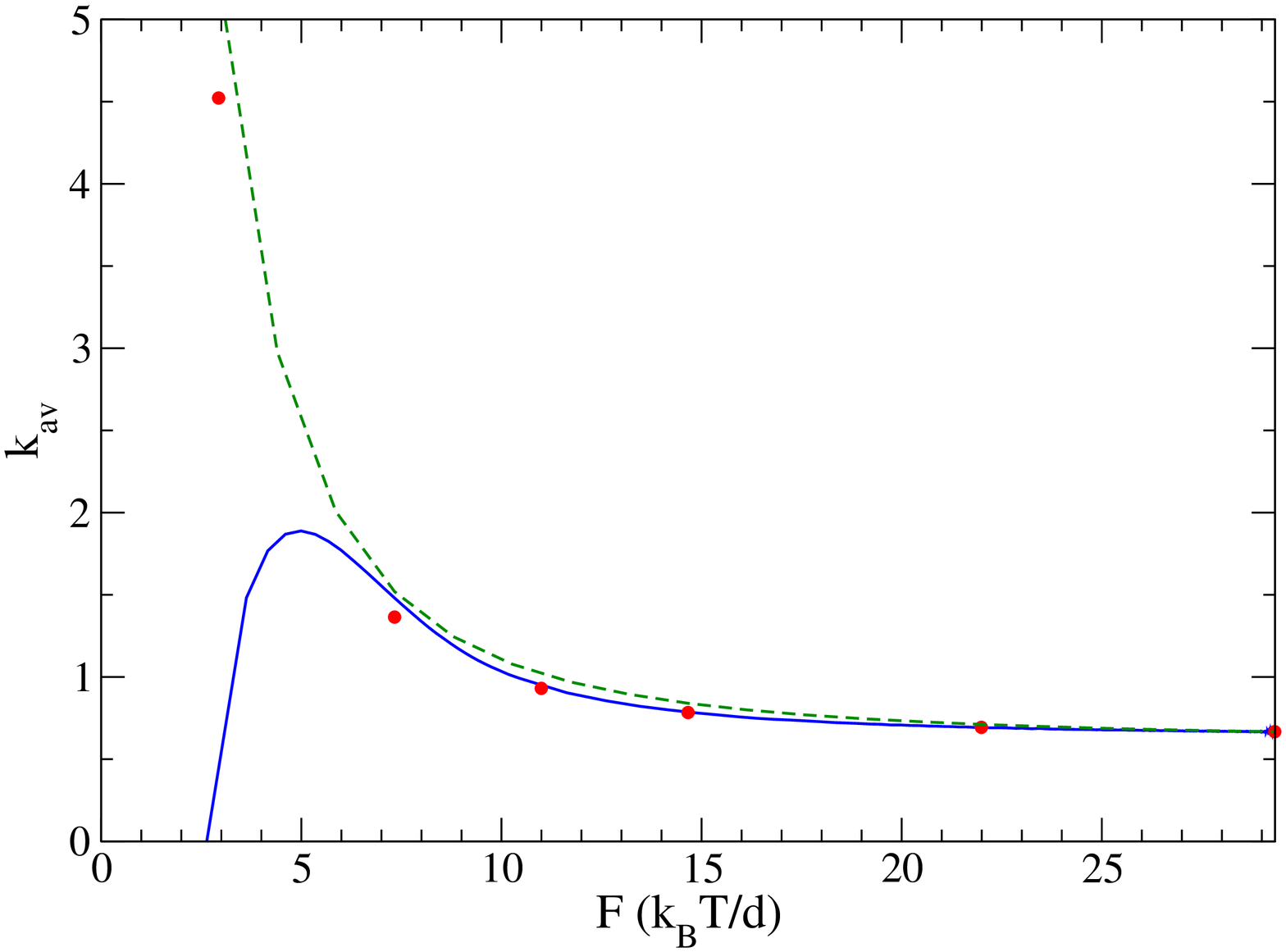}
\caption{Average filament relative size for $N_f=32$ at $\hat{\rho}_1=2.5$ and $\epsilon=0$. The red filled circles, denoting $k_{av}$, are obtained in the stationary regime of constant load runs at each shown load value $F$. Error bars are less than symbol sizes. The dashed green line is the D{\'e}moulin et al. theoretical estimate of $k_{av}(F)$ based on Eqs.(\ref{eq:c4}). The blue continuous curve is obtained for the optical trap set--up by eliminating from $\<k\>_t=\<\sum_{k=0}^{\infty} k \hat{g}(k)\>_t$ and $\<F\>_t=\kappa_T \<L\>_t$ at $\kappa_T=0.4511$, the time parameter $t$.} 
\label{kavdata}
\end{figure}

Figures \ref{g0data} and \ref{kavdata} show respectively, for the stationary state, the load dependent averages $g(0;F)$ and $k_{av}(F)$ Eq.\mref{eq:kn,eq:gk}. D\'emoulin's predictions for the same quantities are also shown in these two Figures, confirming their quantitative accuracy.

\subsection{Optical trap}

Let us start this section with an important remark: for our model, the choice of $\kappa_T$ appears to be completely arbitrary, although, of course, it should satisfy at least the condition that the final equilibrium value of the length of the bundle is much greater than $d$, $\<\hat{L}\>_{EQ}/d\gg1$, in order to avoid boundary effects. 
However, we will see below that this choice will guarantee the equivalence of the results of the optical trap set--up against the constant load, at least for non diverging $\tau_{micro}$. 

\begin{figure}
\centering
\includegraphics[width=0.6\textwidth]{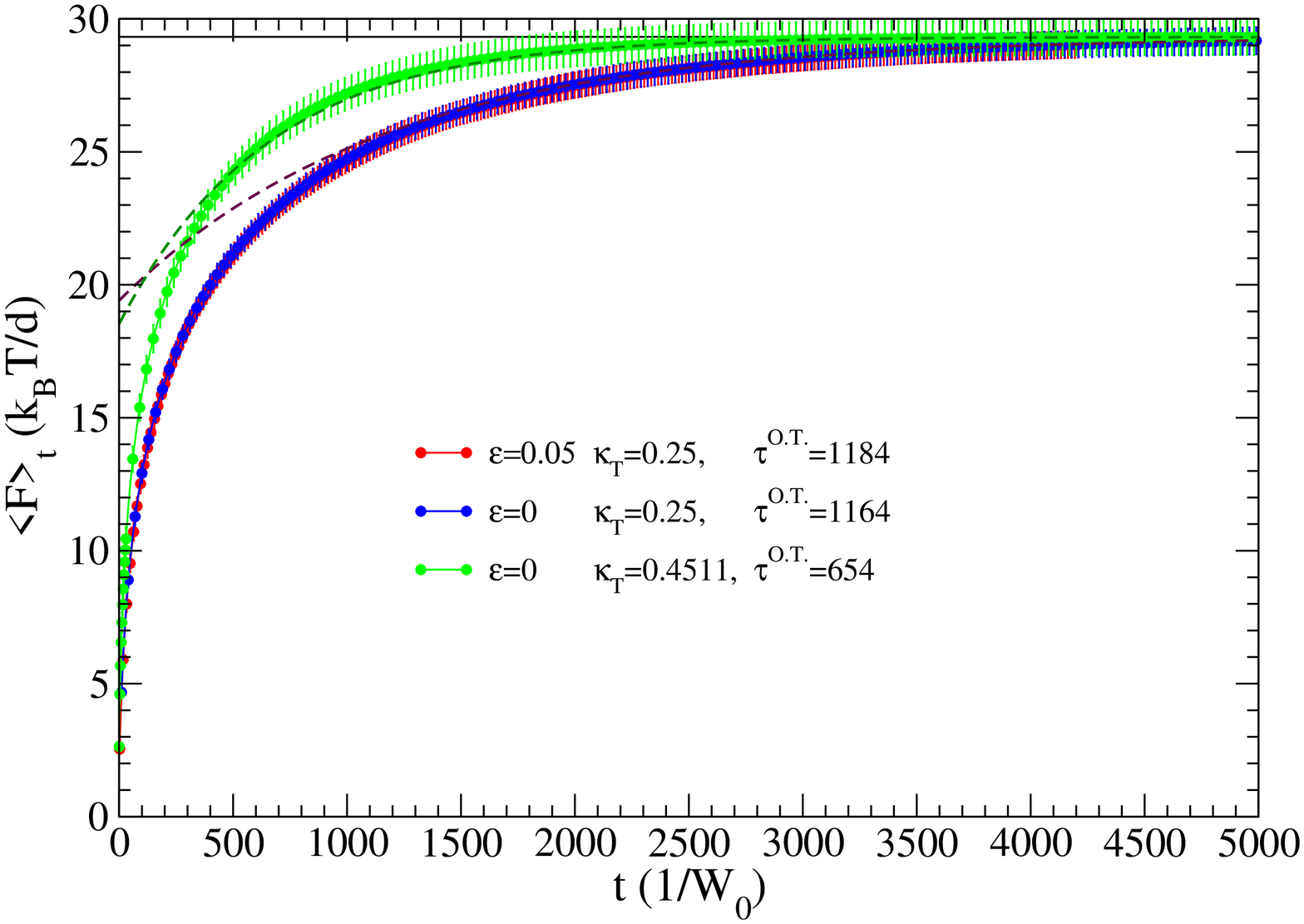}
\caption{Non--equilibrium relaxations of staggered bundles of $N_f =32$ rigid filaments growing in an optical trap at reduced density $\hrho=2.5$, all of them starting from initial conditions with the wall set to a value $L_0 \approx 5d \ll \<L\>_{EQ}$. The wall position $\<L\>_t$ and the associated root mean square deviation $\sigma_L(t)$ are found as a function of time in the figure where what is effectively shown is the load evolution $\kappa_T \<L\>_t$ and corresponding RMSD $\kappa_T \sigma_L(t)$. The final plateau value of the relaxations is compatible with the value $F_s=F_s^H$ given by Eq.\eqref{eq:Fstal} indicated by an horizontal thin black line. The dashed lines represent the best fit of an exponential asymptotic behavior Eq.\eqref{eq:Fasymp}, providing estimates of $\tau^{OT}$ and hence of the chemical friction $\gamma$ defined by Eq.\eqref{eq:taubis}. We find $\gamma=291 \pm 4$ (IWA with $\kappa_T=0.25$), $\gamma=295 \pm 5$ (IWA with $\kappa_T=0.4511$) on the basis of the $\tau^{OT}$ values obtained.}
\label{Fig1}
\end{figure}

Figure  \ref{Fig1} shows time--dependent averages, $F_t=\kappa_T\<\hat{L}\>_t$, for optical trap relaxations computed by EWA and IWA for $\kappa_T=0.25$ and only by IWA for $\kappa_T=0.4511$. In the EWA case, the relaxations start from a bundle size, short with respect to the final equilibrium value, i.e. $L_0=5d$, while in the IWA case the filament sizes all start at $j_n=6$.
The results, obtained by the two algorithms for $\kappa_T=0.25$, are indistinguishable, confirming the validity of the simplified algorithm. The EWA algorithm has been used with a value for $\epsilon$ of $0.05$ that clearly indicates the validity of the simplified computation done in the $\epsilon=0$ limit.
Note that the plateau values give the stalling force predicted by Hill, Eq.\eqref{eq:Fstal}, within statistical error bars. Fluctuations of $\hat{L}$ at equilibrium is given, as expected \cite{Paper2}, by $\sigma_L^{eq}=\sqrt{\left(k_BT/\kappa_T\right)}$ . 
The vertical bars reported in the figure represent the standard deviation, $\kappa_T\sigma_L(t)$, associated to the fluctuation of the force. They remain bounded along the entire curve by the equilibrium value, indicating a limited fluctuation between individual trajectories $\hat{L}(t)$. 
That's relevant because experiments performed by an optical trap set--up usually refer to single trajectory measurement \cite{Dogterom.07}, whose validity is guaranteed by the smallness of fluctuations. 

As Eq.\eqref{FPdiscr} refers to a Markov process, one expects an asymptotic relaxation of $\<L\>_t$ as
\begin{align}
 \<L\>_t=\<L\>_{EQ}+ A_1 \exp{\left(\lambda_1 t \right)}+\dots=\<L\>_{EQ}+ A_1 \exp{\left(- \frac{t}{\tau^{OT}} \right)}+\dots
\label{eq:relax}
\end{align}
where $A_1$ is the amplitude (dependent on initial conditions) of the slowest, non--zero, mode with eigenvalue $\lambda_1=-\frac1{\tau^{OT}}$ of the generator matrix governing the dynamics of the system.
In the same long time limit, one has
\begin{align}
\<F\>_t&=\kappa_T \<L\>_t=F_s + A_1 \kappa_T \exp{\left(\lambda_1 t \right)}+\dots\label{eq:Fasymp}\\
\<V\>_t&= A_1 \lambda_1 \exp{\left(\lambda_1 t \right)}+\dots
\label{eq:Vasymp}
\end{align}
and thus, formally one can express the longest relaxation time of the optical trap relaxation as
\begin{align}
\tau^{OT}=-\lambda_1^{-1}= {\kappa_T}^{-1} \lim_{t\rightarrow \infty}{\frac{F_s-\<F\>_t}{\<V\>_t}}
\label{eq:tau}
\end{align}
From the data in Figure \ref{Fig1} one gets for $\kappa_T=0.25$ and from EWA trajectories $\tau^{OT}=1185 \pm 50$, while from IWA $\tau^{OT}=1164 \pm 10$. For the only IWA case at $\kappa_T=0.4511$ $\tau^{OT}=654 \pm 10$.
By numerical differentiation we have calculated the slopes of $\<L\>_t\equiv l(t;\kappa_T)$, $\<V\>_t=\frac{d\<L\>_t}{dt}\equiv v(t;\kappa_T)$. Eliminating $t$ from the pair of parametric equations $[\<F\>_t=\kappa_T  l(t;\kappa_T), v(t;\kappa_T)]$, we can get the velocity as a function of the force, still a function of $\kappa_T$. The force--velocity relationship for $\kappa_T=0.4511$, shown in Figure \ref{rigid_vF}, turns out to be equivalent, except at small loads, to $V^{stat}(F)$ previously established for the constant force load stationary state. Identical results are obtained from the relaxation with $\kappa_T=0.25$ (not shown), indicating a weak dependence, if any, of the force--velocity relationship on $\kappa_T$.

\subsection{Adiabaticity}

Figure \ref{rigid_vF} shows that, except at low forces (short time part), $V^{stat}(F)$ superposes well the velocity force relationship extracted from the optical trap relaxation, independently of $\kappa_T$. 
The identity between the stationary force--velocity relationship with the one obtained by the relaxation process in the optical trap set--up is a clear indication of the fact that the optical trap set--up is working in adiabatic conditions, i.e. that we have a relaxation process happening in between stationary states. We can derive from this apparent adiabaticity, especially valid at long times when the load changes slowly in time, that 
\begin{align}
\tau^{OT}&\equiv{\kappa_T}^{-1} \lim_{t\rightarrow \infty}{\frac{F_s-\<F\>_t}{\<V\>_t}}
= - \left[{\kappa_T} \left(\frac{\partial V^{stat}}{\partial F}\right)_{F_s}\right]^{-1}\equiv\frac{\gamma}{\kappa_T}
\label{eq:taubis}
\end{align}
where $V^{stat}(F)$ is the constant load force--velocity relationship and where $\gamma$, defined as minus the inverse of the slope of $V^{stat}(F)$ at stalling in Eq.\eqref{eq:taubis}, $\gamma=-\left[\left(\frac{\partial V^{stat}}{\partial F}\right)_{F_s}\right]^{-1}$, is a friction coefficient having a chemical (and not hydrodynamic) origin. The structure of the relaxation time expression Eq.\eqref{eq:taubis} resembles that of an overdamped brownian oscillator.

Eq.\eqref{eq:taubis} can be tested  with our data. Using $\tau^{OT}$ estimates mentioned earlier for the two values of $\kappa_T$, we get three compatible $\gamma$ estimates ($291 \pm 4$ for IWA run at $\kappa_T=0.25$, $296 \pm 12$ for EWA at $\kappa_T=0.25$, and $295 \pm 5$ for IWA at $\kappa_T=0.4511$). 
These values provide an overall estimate of $\gamma=293 \pm 3$ which has to be compared to the value of the slope of $V^{stat}(F)$ at stalling.
The numerical derivative estimated with our too spread data gives $\gamma=272$; unfortunately this value is not sufficiently precise to be completely reliable. Certainly, the uncertainty provided by computing the left and right incremental ratios giving respectively $\gamma=202$ and $\gamma=414$ tells us that we are within the numerical uncertainty. As for the D\'emoulin result, its approximate estimate of the slope leads to $\gamma^{Dem}=281.6$ (see appendix \ref{appC}).

By referring the chemical friction coefficient $\gamma$ to the value characteristic of the mean field force--velocity relationship Eq.\eqref{eq:Fstal}, $\gamma^{MF}=\frac{N_f k_BT}{d^2 W_0}$ we can define a new adimensional coefficient, $C(N_f,\hrho)$, as
\beq
C(N_f,\hrho)=\frac{\gamma}{\gamma^{MF}}=\frac{d^2 W_0}{N_f k_BT}\gamma=9.2\pm0.1 \label{eq:jp}
\eeq
giving a measure of the dynamic correlations between filaments. That means obviously $C=1$ not only in the MF case, but also in the single filament brownian ratchet in the $\epsilon=0$ limit because the force--velocity relationship is identical to the MF expression for $N_f=1$. The D\'emoulin estimate of $C$ gives in our case ($N_f=32,\hrho=2.5$) $C^{Dem}=8.8$. 

On an intuitive basis, adiabadicity is related to a very fast equilibration of filament sizes along the non--equilibrium evolution of the optical trap, with respect to the microscopic relaxation time of the filaments under constant load, $\tau_{micro}$. 
The characteristic time of the optical trap equilibration is $\tau^{OT}$. 
We have seen, in Figure \ref{fig:SM_tmicro}, that, for $F/F_s>0.15$, the typical microscopic relaxation time $\tau_{micro}$ lies in the range $\approx (1\div3) W_0^{-1}$.
Now we can explain what we have anticipated at the beginning of this section: with the values we have chosen for $\kappa_T$, corresponding to equilibrium sizes of the bundle well satisfying the condition $L_{EQ}/d\gg1$ (to avoid boundary problems associated to the short size of the bundle), the values of $\tau^{OT}$ result automatically to be two to three order of magnitude larger (see the values given in Fig.\ref{Fig1}). 
It is important to stress, however, that $\tau_{micro}$ values diverge as $F/F_s\to 0$, a property paralleled by the divergence in the same limit of $k_{av}$.

\section{Concluding remarks}\label{concl_sec}

In this work, we have considered, in a Markovian approximation, a stochastic dynamical model to compute the evolution and the statistical properties of a staggered bundle of $N_f$ rigid living filaments growing against a loaded wall.
In the Fokker--Planck equations we have written down to give an explicit dynamics to our system, a parameter, $\epsilon=\tau_D/\tau_{chem}$, plays a special role.
Generally, the model has to be solved for values of $\epsilon$ relatively small. 
It is found that if we take the $\epsilon=0$ limit, the dynamics simplifies and the overall computations become much lighter.
We have shown numerically that the results obtained in a reasonable range of non--zero values of $\epsilon$ in the neighborhood of zero, coincide with the results obtained using the limiting model and the simplified algorithm.
This indicates the robustness of the $\epsilon=0$ limit.
As a consequence, the major part of the computational work of the present paper has been performed in this limit.
As we have told before, for the loading of the wall, we considered two classical protocols: a constant load or an optical trap set--up, characterized by a harmonic restoring force.
By a series of computer experiments in the case of a constant load and by only one suitable relaxation calculation in the optical trap, we have obtained for the two protocols the classical force--velocity relationship.
With the exception of the region of very weak loads, we have found perfect coincidence of the results.
We have been able to explain this universality of the response of the system as a result of the time scale separation between the relaxation time needed by the wall to adjust to a change of the external force and the characteristic time needed by the chemistry to change the conformation of the bundle.
This condition is violated when the load is very small and that is why the optical trap and constant load results differ, even dramatically, in that region.
Our results suggest that experiments measuring the force--velocity relationship with a harmonic load offer in principle, many advantages over the approach where constant force set--ups are used. Indeed, only a single sample is needed to get a $V(F)$ estimate over a large $F$ window in the first case while a separate experiment and in general a specific sample is needed for each steady state at constant load $F$ investigated. 
Alternative protocols are possible, like imaging techniques used in reference \cite{Dogterom.1997}, but the rules needed to get adiabaticity are easily transposed.
We have been also able to confirm the large scale validity of the approximate theory developed by D\'emoulin et al. \cite{DCBB.14} to compute the properties of our system.

In this work, we have only considered rigid filament. 
Interpreting experimental data with rigid models implies that the semi--flexible character of living biofilaments has limited influence on the results. 
How the bundle dynamics is affected by the flexibility is a delicate point, which is largely unknown and this, to some extent, hampers the confidence in interpreting data with rigid filament models. 
Work is in progress to clarify the influence of flexibility on the force--velocity relation. 

\begin{acknowledgments}
We thank G. Kozyreff for his help with the perturbation expansion of the F.P. equation in section 2B. Two of us (CP and JPR) are grateful to J. Baudry, J.F. Joanny  et D. Lacoste for useful discussions. We thank G. Destr\'ee for technical help. JPR thanks the University of L'Aquila for hospitality during a three months visit. 
CP is supported by the Agence Nationale de la Recherche (ANR) under the project ``HyLightExtreme''.
AP is supported by a Mobility Grant for PhD students from Sapienza University of Rome, and thanks the ENS for hospitality during a six months visit.
\end{acknowledgments}

\appendix
\section{Discretized Fokker-Planck Equation for the wall--bundle system in an optical trap or constant force set--up}\label{appA}

In this appendix we derive a proper discretization of the Fokker--Planck equation together with the elements of the generator matrix $\mathbf{Q}$ of the Markov process given by Eq.\eqref{FPdiscr}.

To get the matrix elements which account for the discretization of the variable $L$, following the procedure introduced in \cite{Wang2003}, we concentrate only on the diffusive part of Eq.\eqref{FPeq} for the wall position probability at given chemical state, $P_{j_1,\dots,j_{N_f}}(L,t)\equiv P_j(L,t)$:
\begin{equation}
\frac{\partial P_j(L,t)}{\partial t}=-\frac{\partial}{\partial L}J_j(L,t)
\label{eqA1}
\end{equation}
where $J_j(L,t)=-D\left(\frac{\partial P_j(L,t)}{\partial L}+\frac1{k_BT}\frac{d\Phi}{dL}P_j(L,t)\right)$ is the probability current, with $\frac{d\Phi}{dL}=\kappa_TL$ or $-F$ for respectively the optical trap or the constant force set--up.
We define the probabilities for the wall to be in the intervals ($l=L/\delta$) $k-1/2\leqslant l<k+1/2$ and $k+1/2\leqslant l<k+3/2$ as:
\bea
p_k(t)&=&\int_{k-1/2}^{k+1/2}P_j(l,t)dl\label{eqA2_1}\\
p_{k+1}(t)&=&\int_{k+1/2}^{k+3/2}P_j(l,t)dl.
\label{eqA2_2}
\eea
By defining the wall \textit{forward rate} $F_{k+1/2}$ of going from $k$ to $k+1$ ($F_{k-1/2}$ from $k-1$ to $k$) and the wall \textit{backward rate} $B_{k+1/2}$ of going from $k+1$ to $k$ ($B_{k-1/2}$ from $k$ to $k-1$), the time evolution of the probability $p_k(t)$ can be written as \cite{Wang2003}:
\bea
\frac{dp_k(t)}{dt}&=&F_{k-1/2}p_{k-1}-(F_{k+1/2}+B_{k-1/2})p_k+B_{k+1/2}p_{k+1}\nonumber\\
&=&-(F_{k+1/2}p_k-B_{k+1/2}p_{k+1})+(F_{k-1/2}p_{k-1}-B_{k-1/2}p_k)\nonumber\\
&=&-(J_{k+1/2}-J_{k-1/2})
\label{eqA3}
\eea
where the rates $F_{k\pm1/2}$ and $B_{k\pm1/2}$ have to be derived by discretizing Eq.\eqref{eqA1}. $J_{k+1/2}$ is the net probability flux between sites $k$ and $k+1$ ($J_{k-1/2}$ is between $k-1$ and $k$).

If we now discretize Eq.\eqref{eqA1} using e.g. the central difference method ($f'_{k+1/2}=(f_{k+1}-f_k)/\delta $) and compare the resulting discrete equation with Eq.\eqref{eqA3}, the forward and backward rates obtained will not respect the detailed balance, a sufficient condition to reach equilibrium, while we expect the evolution of the Markov chain to lead to it, with each process balanced by its reverse. 

To overcome this difficulty, following \cite{Wang2003}, we can look for the stationary solution of Eq.\eqref{eqA1} and see if, by integration over a proper interval of lengths, we can obtain an identification of the rates, bringing us to coefficients satisfying the detailed balance.

Looking at the definitions Eqs.\mref{eqA2_1,eqA2_2}, we see that to get $p_k$ and $p_{k+1}$ from a solution of the stationary equation \eqref{eqA1} we need to solve it in the interval $(k-1/2,k+3/2)$. Then we look for the solution of the probbaility $P_j(L,t)$ in terms of the stationary solution $P_{EQ}(l)$ of:
\beq
D\frac{d}{dl}\left(\frac{dP_{EQ}(l)}{dl}+\frac{\Delta\Phi_{k+1/2}}{k_BT}P_{EQ}(l)\right)=0\qquad 
\label{FPdiff}
\eeq
in $l\in\left(k-1/2,k+3/2\right)$, where we have substituted to $d\Phi/dl$ by the constant approximation $\Delta\Phi_{k+1/2}$, with
\beq
\Delta\Phi_{k+1/2}=\Phi(k+1)-\Phi(k)
\eeq 
The general solution of Eq.\eqref{FPdiff} is $P_{EQ}(l)=\eta\exp\left(-\frac{\Delta\Phi_{k+1/2}}{k_BT}l\right)+\theta$ with $\eta$ and $\theta$ constants. Plugging this expression into Eqs.\mref{eqA2_1,eqA2_2}, one can easily find $\eta$ and $\theta$ in terms of $p_k$ and $p_{k+1}$.
Then the (approximate) stationary solution of the Fokker--Planck equation for the wall in the interval $(k-1/2,k+3/2)$ is:
\bea
P_{EQ}(l)&=&\frac{\Delta\Phi_{k+1/2}\(p_k-p_{k+1}\)}{k_BT\(\exp\(-\frac{\Delta\Phi_{k+1/2}}{k_BT}\)-1\)^2}\exp\(\frac{\Delta\Phi_{k+1/2}}{k_BT}\(k-1/2\)\)\exp\(-\frac{\Delta\Phi_{k+1/2}}{k_BT}l\)\nonumber\\
&+&\frac{p_k\exp\(-\frac{\Delta\Phi_{k+1/2}}{k_BT}\)-p_{k+1}}{\(\exp\(-\frac{\Delta\Phi_{k+1/2}}{k_BT}\)-1\)}\qquad\qquad\qquad l\in\(k-1/2,k+3/2\)
\eea
From this equation we get the probability flux in the same interval:
\bea
J_{EQ}(l)&=&-\widetilde{D}\frac{dP_{EQ}(l)}{dl}-\widetilde{D}\frac{\Delta\Phi_{k+1/2}}{k_BT}P_{EQ}(l)=-\frac{\widetilde{D}\Delta\Phi_{k+1/2}}{k_BT}\frac{p_k\exp\(-\frac{\Delta\Phi_{k+1/2}}{k_BT}\)-p_{k+1}}{\left(\exp\left(-\frac{\Delta\Phi_{k+1/2}}{k_BT}\right)-1\right)}
\eea
with $\widetilde{D}=D/\delta^2$ the diffusion constant in $\delta$ units.
Comparing this current with the probability flux defined in Eq.\eqref{eqA3}, we get the following forward and backward rates:
\bea
\label{forward}F_{k+1/2}&=&\widetilde{D}\frac{\Delta\Phi_{k+1/2}/k_BT}{\exp\(\frac{\Delta\Phi_{k+1/2}}{k_BT}\)-1}\\
\label{backward}B_{k+1/2}&=&\widetilde{D}\frac{-\Delta\Phi_{k+1/2}/k_BT}{\exp\(-\frac{\Delta\Phi_{k+1/2}}{k_BT}\)-1}.
\eea

The same approach for the interval $(k-3/2,k+1/2)$ can be used to get $F_{k-1/2}$ and $B_{k-1/2}$.

By direct substitution, we see that Eqs.\mref{forward,backward} respect the detailed balance, $F_{k+1/2}P_{EQ}(k)=B_{k+1/2}P_{EQ}(k+1)$.
Substituting the appropriate expression for $\Phi(L_k)$, we have:
\beq
\Delta\Phi_{k+1/2}=\begin{cases}F\delta&\qquad \mbox{constant load}\\
\frac12\kappa_T\delta^2\((k+1)^2-k^2\)&\qquad \mbox{optical trap}
\end{cases}
\label{en_diff}
\eeq

The non--zero elements of the generator matrix $\mathbf{Q}$ can now be written as follows:
\bea
&&\mathcal{Q}_{\{j_1,\dots,j_n,\dots,j_{N_f},k\}\{j_1,\dots,j_n,\dots,j_{N_f},k+1\}}=F_{k+1/2} \\
&&\mathcal{Q}_{\{j_1,\dots,j_n,\dots,j_{N_f},k\}\{j_1,\dots,j_n,\dots,j_{N_f},k-1\}}=C_{k-1/2}=\begin{cases} B_{k-1/2} \qquad&\text{if }k-1\geqslant X^\ast/d\\
0\qquad&\text{otherwise}
\end{cases}
\label{backw2}\\
&&\mathcal{Q}_{\{j_1,\dots,j_n,\dots,j_{N_f},k\}\{j_1,\dots,j_n+1,\dots,j_{N_f},k\}}=U_{j_n}=\begin {cases}  U_0 \qquad&\text{if }k\geqslant X_n(j_n+1)/d\\
0\qquad&\text{otherwise}
\end{cases}\\
&&\mathcal{Q}_{\{j_1,\dots,j_n,\dots,j_{N_f},k\}\{j_1,\dots,j_n-1,\dots,j_{N_f},k\}}=W_{j_n}=W_0 \\
&&\mathcal{Q}_{\{j_1,\dots,j_n,\dots,j_{N_f},k\}\{j_1,\dots,j_n,\dots,j_{N_f},k\}}=-F_{k+1/2}-C_{k-1/2}-\sum_{n=1}^{N_f}\left(U_{j_n}+ W_{j_n}\right)\label{diagonal1}
\label{eq:dmcF}
\eea
with $X_n(j_n)$ and $X^\ast$ given by Eqs.\mref{Xn,Xstar}.
The row sums of this matrix are zero, as required for a generator matrix of a Markov chain:
\beq
\sum_{\{j_1',\dots,j_n',\dots,j_{N_f}',k'\}}\mathcal{Q}_{\{j_1,\dots,j_n,\dots,j_{N_f},k\}\{j_1',\dots,j_n',\dots,j_{N_f}',k'\}}=0.
\eeq
Eq.\eqref{FPdiscr} represents hence a continuous time Markov process with discrete states; as for the variable $L$, the discrete states are approximations (exact in the $\delta\to0$ limit) to the continuous/discrete process defined in Eq.\eqref{FPeq}. 

\section{Elements of the $\epsilon=0$ generator matrix}\label{appB}

In this appendix we write explicitely the matrix elements of $\mathcal{Q}^{(0)}$, generator of the Markov process in the $\epsilon=0$ limit Eq.\eqref{CTMC0}.
Since in this limit the integration in $L$ allowed us to get rid of the continuous wall diffusion process, these elements can be written immediately:
\bea
&&\mathcal{Q}^{(0)}_{\{j_1,\dots,j_n,\dots,j_{N_f}\}\{j_1,\dots,j_n+1,\dots,j_{N_f}\}}=U_{j_n}=U_0A^{(n)}(j_1,\dots,j_n,\dots,j_{N_f})\\
&&\mathcal{Q}^{(0)}_{\{j_1,\dots,j_n,\dots,j_{N_f}\}\{j_1,\dots,j_n-1,\dots,j_{N_f}\}}=W_{j_n}=W_0 \\
&&\mathcal{Q}^{(0)}_{\{j_1,\dots,j_n,\dots,j_{N_f}\}\{j_1,\dots,j_n,\dots,j_{N_f}\}}=-\sum_{n=1}^{N_f}\left(U_{j_n}+ W_{j_n}\right)
\eea
where $A^{(n)}(j_1,\dots,j_n,\dots,j_{N_f})$ is given in by:
\bea
A^{(n)}(j_1,\dots,j_n,\dots,j_{N_f})&=&\int_{X^\ast/d}^\infty dx\;\Theta\left(x-X_n(j_n+1)/d\right)\widetilde{P}_{EQ}(x\;|\; j_1,\dots,j_n,\dots,j_{N_f})\nonumber\\
&=&\begin{dcases*}
\exp\left[-f\left(X^{*'}-X^*\right)/d\right]\qquad &\mbox{constant load}\\
\frac{\erfc\left[\left(\widetilde{\kappa}_T/2\right)^{1/2}X^{*'}/d\right]}{\erfc\left[\left(\widetilde{\kappa}_T/2\right)^{1/2}X^*/d\right]}\qquad &\mbox{optical trap}
\end{dcases*}
\label{An}
\eea
where $X^{*'}$ is the most advanced filament's tip for the set of filament sizes $\{j_1,\dots,j_n+1,\dots,j_{N_f}\}$.
Eq.\eqref{An} has been derived previously for constant load \cite{vanDoorn2000,DCBB.14} and for optical trap load \cite{Carlsson}.

\section{D\'emoulin et al. prediction for $V(F)$ and $k_{av}$}\label{appC}
D\'emoulin et al. \cite{DCBB.14} have proposed an approximate solution for the force--velocity  relationship of staggered rigid filaments subjected to a constant load $F$ in the $\epsilon=0$ limit.
They found:
\bea
V(F)&=&\frac{dU_0}{N_f}\left[N_f\exp\left(-\frac{Fd}{k_BT}\right)+\sum_{m=1}^{N_f-1}g(0)\left(N_f-m\right)\exp\left(-\frac{Fd(N_f-m)}{N_fk_BT}\right)\right]\nonumber\\
&-&\frac{dW_0}{N_f}\left[g(0)\sum_{m=1}^{N_f-1}m\left(1-g(0)\right)^{m-1}+N_f\left(1-g(0)\right)^{N_f-1}\right]
\label{vf_dem}
\eea
with the relative size distribution $g(k)$ given by:
\beq
g(k)=\frac{d(U_0-W_0)-V}{dU_0}\left(\frac{V+dW_0}{dU_0}\right)^k\qquad k=0,\infty.
\label{g_k}
\eeq
It can be verified that at stalling, $F=F_s=N_f\frac{k_BT}{d}\ln\hrho$, one gets $V=0$ and $g(0)=1-\hrho^{-1}$.

For the comparison in the text, we need to compute $k_{av}$ and $V(F)$ explicitely:
\begin{enumerate}
\item $\mathbf{k_{av}}$: defining $\xi=\frac{V+dW_0}{dU_0}$ to simplify expressions, one gets from Eq.\eqref{g_k}:
\bea
g(k)&=&(1-\xi)\xi^k\label{eq:c3}\\
k_{av}&=&\sum_{k=1}^\infty kg(k)=\frac\xi{1-\xi}\label{eq:c4}
\eea
\item $\mathbf{V(F)}$: Inserting $g(0)=\frac{d(U_0-W_0)-V}{dU_0}$ into Eq.\eqref{vf_dem}, we find for $V(F)$ a polynomial equation in $V$ to solve. 
Writing $v=V/dW_0$, we find:
\bea
\phi(v)&=&\frac\hrho{N_f}\left[N_f\exp\left(-\frac{Fd}{k_BT}\right)+\sum_{m=1}^{N_f-1}\left(1-\frac{1+v}\hrho\right)\left(N_f-m\right)\exp\left(-\frac{Fd(N_f-m)}{N_fk_BT}\right)\right]\nonumber\\
&-&\frac{1}{N_f}\left[\left(1-\frac{1+v}\hrho\right)\sum_{m=1}^{N_f-1}m\left(\frac{1+v}\hrho\right)^{m-1}+N_f\left(\frac{1+v}\hrho\right)^{N_f-1}\right]-v=0
\label{vf_dem1}
\eea

Eq.\eqref{vf_dem1} can be solved numerically using the Newton--Raphson method, for which the derivative of $\phi(v)$ with respect to $v$ is needed:
\bea
\phi^\prime(v)&=&-\frac\hrho{N_f}\left[\sum_{m=1}^{N_f-1}\left(N_f-m\right)\left(N_f-m\right)\exp\left(-\frac{Fd(N_f-m)}{N_fk_BT}\right)\right]\\
&+&\frac{1}{N_f\hrho}\left[\sum_{m=1}^{N_f-1}m\left(\frac{1+v}\hrho\right)^{m-2}\left(\frac{1+v}\hrho-\left(1-\frac{1+v}\hrho\right)(m-1)\right)-N_f\left(N_f-1\right)\left(-\frac{1+v}\hrho\right)^{N_f-2}\right]-1\nonumber
\eea
The Newton--Raphson method requires a first guess value, say $v_0$, which can be taken as e.g. the Hill's value.
\end{enumerate}
The solution of D\'emoulin et al. equation for a bundle of $N_f=32$ filaments and supercritical density $\hrho=2.5$ is reported in Fig.\ref{Fig1}, to be compared with the results of our stochastic dynamics algorithm.
The same is done, substituting $V(F)$ in Eqs.\mref{eq:c3,eq:c4}, for $g(0)$ and $k_{av}$ in Fig.\ref{g0data} and \ref{kavdata} respectively. 

To predict within the present theory the value of $\tau^{OT}$, we need to compute the derivative of $V(F)$ with respect to $F$ at stalling, obtaining $\gamma^{Dem}$. From Eq.\eqref{vf_dem}, we get $dV/dF$ as an implicit function of $V(F)$ and $F$. At stalling $F=F_s$, $V(F_s)=0$, we obtain:
\beq
\frac{\partial V}{\partial F}\Bigg|_{F=F_s}=-\frac{d^2W_0}{k_BT}\frac{\hrho^{1-N_f}\left[1+\sum_{m=1}^{N_f-1}\hrho^m(1-\hrho^{-1}\frac{(N_f-m)^2}{N_f^2}\right]}{1+N_f^{-1}\sum_{m=1}^{N_f-1}m\hrho^{-m}+(N_f-1)\hrho^{-(N_f-1)}-N_f^{-1}\sum_{m=1}^{N_f-1}m\hrho^{-m}\left[1-(\hrho-1)(m-1)\right]} \label{eq:der}
\eeq
For our conditions the value is $\gamma^{Dem}=-(dV/dF)_s^{-1}=281.6$, in agreement with our results (see main text).

\end{document}